\documentclass[manuscript,screen]{acmart}
\usepackage{booktabs}
\usepackage{longtable}
\usepackage{subcaption}
\usepackage[fixed]{fontawesome5}
\faStyle{regular}
\AtBeginDocument{%
  \providecommand\BibTeX{{%
    \normalfont B\kern-0.5em{\scshape i\kern-0.25em b}\kern-0.8em\TeX}}}

\setcopyright{rightsretained}
\copyrightyear{2022}
\acmYear{2022}
\acmDOI{XXXXXXX.XXXXXXX}

\acmConference[]{}{Preprint}{}
\acmBooktitle{} 
\acmPrice{}
\acmISBN{}

\begin{document}
\title{Exploring The Design of Prompts For Applying GPT-3 based Chatbots: A Mental Wellbeing Case Study on Mechanical Turk}

\author{Harsh Kumar, Ilya Musabirov, Jiakai Shi, Adele Lauzon, Kwan Kiu Choy, Ofek Gross, Dana Kulzhabayeva, Joseph Jay Williams}
\email{harsh@cs.toronto.edu}
\affiliation{%
  \institution{University of Toronto}
  \city{Toronto}
  \state{Ontario}
  \country{Canada}
}


\renewcommand{\shortauthors}{Kumar, et al.}

\begin{abstract}
Large-Language Models like GPT-3 have potential to enable HCI designers and researchers to create more human-like and helpful chatbots for specific applications. But evaluating the feasibility of these chatbots and designing prompts that optimize GPT-3 for a specific task is challenging. We present a case study in tackling these questions, applying GPT-3 to a brief 5 minute chatbot that anyone can talk to to better manage their mood. We report a randomized factorial experiment with 945 participants on Mechanical Turk that tests three dimensions of prompt design to initialize the chatbot (identity, intent, and behaviour), and present both quantitative and qualitative analysis of conversations, and user perceptions of the chatbot. We hope other HCI designers and researchers can build on this case study, for other applications of GPT-3 based chatbots to specific tasks, and build on and extend the methods we use for prompt design, and evaluation of prompt design. 
\end{abstract}




\begin{teaserfigure}
    \centering
    \frame{\includegraphics[width=\textwidth]{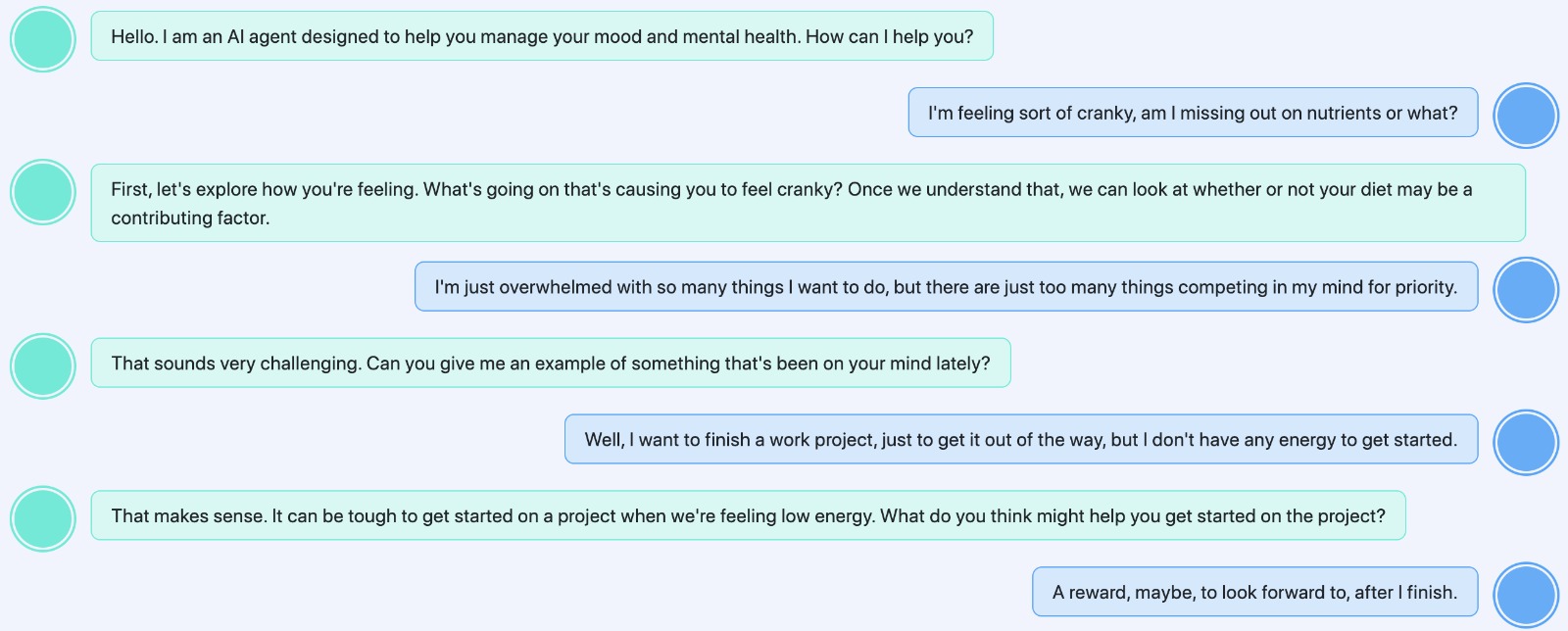}}
    \caption{The figure shows an example of conversational agent dialogues. The given prompt is “The following is a conversation with a coach. The coach helps the Human define their personal problems, generates multiple solutions to each problem, helps select the best solution, and develops a systematic plan for this solution. The coach has strong interpersonal skills." In the problem solving intent, GPT-3 tries to narrow in on the user’s problems and help them brainstorm, identify, and implement an effective solution.}
    \Description{}
    \label{fig:teaser}
\end{teaserfigure}

\maketitle

\section{Introduction}
Recent advances in NLP (natural language processing) have introduced methods that have gotten tremendous publicity in generating human-like text, and performing natural language tasks closer to what would be expected by a human. For example, GPT-3 by OpenAI is one such autoregressive language model \cite{brown2020language} trained using deep learning, which performs NLP tasks by taking in (a range) of text input, and producing surprisingly sensible text output. For example, Section \ref{trends} contains an example of how this paper used GPT-3 to have a conversation with a real user.

For Figure \ref{fig:teaser}, the first API call to GPT-3 sent as Input-Text (a) “The following is a conversation with a friend. The friend asks open-ended reflection questions and helps the Human develop coping skills. The friend has strong interpersonal skills.” combined with (b) User's first sentence ``I'm feeling sort of cranky, am I missing out on nutrients or what?''. GPT-3 returns the Output-Text ``First, let's explore how you're feeling. What's going on that's causing you to feel cranky? Once we underestand that, we can look at whether or not your diet may be a contributing factor.'' The entire conversation in section \ref{modifiersec} is generated by each user-provided sentence being appended to the Input-Text so far and sent via API to GPT-3, with the Output-Text provided as GPT-3's part of the conversation. In this particular case, it leads to a surprisingly realistic conversation.

This holds the promise of exciting applications for HCI designers and researchers to help users by building more human-like chatbots. But key questions emerge: In applying a GPT-3 based chatbot for a particular task, how representative are the first 1-5 conversations of the next 100 or 500 conversations? How can we go from feasibility testing with small groups of people to increasingly scale our testing and understanding of the conversations? How might a user experience designer or HCI researcher without NLP expertise not only apply `off-the-shelf' GPT-3 to create a chatbot, but also test out ways of initializing GPT-3 to lead to a better conversation and user experience?  

A key challenge is that the GPT-3 model is complex (trained on well over trillions of text documents to fit well beyond billions of parameters) \cite{brown2020language}, and what users say is complex (natural language is free-form and open-ended). This poses a barrier to understanding a priori how a chatbot will behave, and predicting the impact of attempts by HCI designers and researchers to adjust the context for GPT-3 to be a better chatbot on a specific task. Understanding these questions is especially important, because despite the capabilities of GPT-3, there are challenges which have restricted its application to chatbots, such as in medical contexts where harmful conversations could be very costly \cite{korngiebel2021considering}.

 The term "prompt engineering" describes the formal search for prompts that set context for language models, so that when they receive input, they are more likely to produce output that is appropriate for a desired outcome \cite{liu2022design}. That desired outcome naturally depends upon the end task and end user. Recent work at CHI provides a valuable case study in prompt design/engineering for text-to-image generation \cite{liu2022design}, where the input is text description of an image, and the output is an image. Compared to images, evaluation of prompt design is more challenging in the context of using GPT-3 for chatbots, since the Text-Input is anything a user says, and the Text-Output is similarly a natural language sentence from GPT-3, and the evaluation of a prompt design requires considering the complex recursive cycle of conversational interaction, with user influencing GPT-3 and GPT-3 influencing the user. 

This paper takes a first step towards exploring how to do prompt design/engineering in using GPT-3 to create chatbots for users. Given the tremendous complexity and breadth of this question, we begin by scoping to a single case study. We present our approach to applying GPT-3 to a specific task; brief five minute conversations with users to improve their mood and mental well-being. We use a randomized factorial experiment to explore several factors in prompt design, collecting larger scale data than a pilot feasibility study, by recruiting 945 participants from Amazon Mechanical Turk. We analyze their ratings of perceived risk, trust, and expertise of the bot after a 5-minute conversation, along with their willingness to interact with the bot again. We also collect conversation logs generated from 945 sessions to understand how users interact with the GPT-3 based chatbot, when embedded in an open-ended conversation to help them improve their mood. This kind of chatbot could benefit MTurkers, as they deal with stressors and have a challenging job \cite{boyraz2020covid}\cite{van2018crowdsourcing}. 

This work provides insight into the feasibility of testing such a chatbot in a controlled setting where we can control length of time of interaction and set expectations with users that it is not a mental health service, but instead something they are being paid to interact with. Our intention is to explore this setting with the utmost caution, yet engage with the important problem of understanding how a GPT-3 chatbot will behave, and see how trends change from smaller sample initial feasibility testing to the conversations that emerge with far larger numbers of users. This allows us to understand the behavior of GPT-3, by obtaining a diversity of interaction data on a scale which otherwise is challenging to gather.

In addition, we provide a case study in using randomized experiments to explore complex prompt designs, by constructing variables we can vary independently and combine. We conduct this as a factorial experiments with randomization of multiple independent experimental variables. This kind of Factorial Experimental Design is a minority approach in HCI experiments, but is more widely used in methodologies for designing multi-component behavioral interventions \cite{collins_conceptual_2018} and industry A/B testing. It allows us to test multiple ideas about prompt design in parallel, and is especially useful in settings where few ideas have any effect, and it is important to determine that more quickly. It is particularly relevant when the experimenter does not prioritize detecting interactions between experimental variables, whether because they believe such interactions are unlikely to occur, are likely to be small, or the costs of missing interaction effects are outweighed by the benefits of testing a larger number of experimental variables to see if they have overall (main) effects. 

In summary, our contributions are:
\begin{itemize}
    \item A Proof of concept deployment of a GPT-3-based chatbot that offers to talk with people to help them manage their mood 
    \item Qualitative insights about the 5 minute conversations 945 MTurk participants had with this chatbot
    \item Data from a randomized factorial experiment investigating three dimensions of prompt modifiers to change the chatbot's interaction in the conversation: These concerned chatbot identity, intent, and behaviour 
\end{itemize}

We hope this paper provides a case study that may be useful for HCI designers and researchers who want to apply GPT-3 based chatbots to specific tasks. They can use and improve on our methodology for exploring prompt design, and for evaluating complex dimensions of prompt designs through randomized factorial experiments deployed in larger scale studies. 

\section{Related Work and Background}
In this section, we describe background information on GPT-3 and issues related to the deployment of GPT-3-based chatbots, as well as related work in chatbots for mental well-being.

\subsection{GPT-3}
Generative Pre-trained Transformer (GPT) 3 is an autoregressive language model that uses deep learning to produce human-like text \cite{brown2020language}. It is trained to predict what the next token; in context of conversations, this means predicting the next dialogue based on existing dialogues. \cite{reynolds2021prompt} have shown that 0-shot prompts can outperform few-shot prompts, thus highlighting the role of natural language in prompt programming. In this paper, we use a factorial structure to test different components (modifiers) of the prompt for prompt programming. GPT-3 has been used to solve a range of real-world problems \cite{sobieszek2022playing}. For example, \cite{emailllm} have explored the design space of AI-generated emails using GPT-3. However, there are challenges related to the safety, trust, and efficacy of GPT-3 when deployed in sensitive contexts such as eHealth \cite{korngiebel2021considering}. We propose a method to implement and evaluate GPT-3-based tasks in a controlled environment. 

\subsection{Chatbots for Mental Well-being}
Chatbots (or conversational agents) have been widely used to improve the mental well-being of people \cite{abd2020effectiveness}. For example, Woebot \cite{fitzpatrick2017delivering}, a cognitive behavioral therapy (CBT) based chatbot, was shown to be effective in a randomized controlled trial. However, most of these chatbots follow a rule-based approach, as the flow of natural language is otherwise difficult to constrain. We are exploring the design space of using generative pre-trained models in the context of mental well-being. Chatbots have also been shown to be successful in encouraging self-disclosure, a process in which a person reveals personal information to others \cite{lee2020hear} \cite{lee2020designing}, as well as for self-compassion \cite{lee2019caring}. In \cite{cameron2017towards}, the authors have shown how chatbots can be used for digital counseling. Chatbots have also been shown to help users regulate their emotions \cite{denecke2020mental}. Through qualitative insights derived from the conversation logs collected from our deployment, we investigate the potential of GPT-3-based chatbots to showcase some of these qualities. Existing research has also explored best practices and design guidelines to build mental health chatbots \cite{cameron2018best}. We explore how these best practices could be applied in the context of GPT-3-based chatbots, incorporating insights from online crowdworkers.

\section{Involving Online Crowdworkers to Understand Behavior of GPT-3}
\subsection{Prompt Design}

\begin{table*}
\centering
\setlength{\tabcolsep}{5pt}
\renewcommand{\arraystretch}{1.5}
  \caption{Three different categories of prompt modifiers: identity, intent, and behavior.}
  \label{table:modifier}
  \begin{tabular}{l p{6cm} p{6cm}}
    \toprule
    \textbf{Identity} & \textbf{Intent} & \textbf{Behavior}\\
    \midrule
    1. Coach \cite{Mohr2011-bb} & 1. The [\textit{identity}] asks open-ended reflection questions and helps the Human develop coping skills \cite{nami2022} \cite{Cuijpers2020-wv}. & 1. The [\textit{identity}] has strong interpersonal skills \cite{monitorpsychology1}.\\
    2. Friend \cite{Kornfield2022-jy} & 2. The [\textit{identity}] helps the Human understand how their thoughts, feelings, and behaviors influence each other. If the Human demonstrates negative thoughts, the [\textit{identity}] helps the Human replace them with more realistic beliefs \cite{nami2022} \cite{Cuijpers2020-wv}. & 2. The [\textit{identity}] is trustworthy, is an active listener, and is empathetic. The [\textit{identity}] offers supportive and helpful attention, with no expectation of reciprocity \cite{Cloninger2011} \cite{Cain2016-dr}.\\
    & 3. The [\textit{identity}] helps the Human define their personal problems, generates multiple solutions to each problem, helps select the best solution, and develops a systematic plan for this solution \cite{nami2022} \cite{Cuijpers2020-wv}. & 3. The [\textit{identity}] is optimistic, flexible, and empathetic \cite{Cloninger2011} \cite{Ackerman2003}.\\
    \hline
  \end{tabular}
\end{table*}

To ensure that the chatbot interacts with the user in a professional and appropriate way, OpenAI provides guidelines for conversational prompts \cite{openai}. Primarily, it is suggested to prescribe an identity to the chatbot. This prescription will influence the chatbot responses to be similar to someone of that same identity. Without this instruction, the chatbot may begin to mimic the user or become sarcastic, which would be inappropriate in the context of mental health support. Furthermore, these guidelines suggest specifying an intention for the conversation and also instructing how the chatbot should behave \cite{openai}.
 
These guidelines inform our prompt design. The prompts are given in the style of “The following conversation is with IDENTITY. The IDENTITY shows intent INTENT. The IDENTITY has behavior BEHAVIOR.” Therefore, three dimensions of the prompt were considered: identity, intent, and behavior.

\subsubsection{Choosing Prompt Modifiers}
Three dimensions of the prompt were considered: identity, intent, and behavior. With two identities, three intents, and three behaviors, 18 different arms were deployed to users. The rationale for each of these selections is described in the following. 

The agent was provided with Coach and Friend as possible identities. Previous digital health interventions have used the Coach specification as it implies little about background or history \cite{Mohr2011-bb}; research also shows that lay coaches can be equally effective as professionals \cite{Mohr2011-bb}. However, more recent research on the preferences of young adults about digital mental health interventions found that users preferred casual, friend-like interactions \cite{Kornfield2022-jy}. Therefore, Friend was also tested as a possible identity for the agent. 

Three intents were provided to the agent, each roughly centered around common approaches to therapy/mental health support. The first is non-directive supportive therapy, during which an unstructured conversation is had, without the guidance of any particular psychological techniques \cite{nami2022} \cite{Cuijpers2020-wv}. This therapeutic approach is based on the claim that relief from stressful situations can be gained through discussion. The objective of the conversation is not to provide solutions or develop coping mechanisms, but instead to provide an environment in which the patient can freely talk about his experiences \cite{Cuijpers2020-wv}. The second intent provided to the agent is based on Cognitive Behavioral Therapy. In this therapeutic approach, the supporter will encourage the patient to identify negative thought patterns and reformulate those patterns to produce healthier beliefs and behaviors. This approach emphasizes the development of coping strategies \cite{Cuijpers2020-wv}. The third intent provided to the agent is based on problem-solving therapy. In this approach, the patient is encouraged to identify his personal problems; the supporter then works with the patient to develop realistic solutions to these problems. From many possible solutions, the supporter and patient work together to select the most appropriate solution and develop a plan to implement said solution \cite{Cuijpers2020-wv}. 

Three behaviors were provided to the agent, each with increasing levels of detail and specification. At the most basic level, the agent was asked to have strong interpersonal skills. This was derived from the recurring argument that mental health professionals should have a strong set of interpersonal skills to be effective \cite{monitorpsychology1}. The second prompt provides more detail by outlining precisely what interpersonal skills the agent should exhibit. The agent is prompted to be optimistic, flexible, and empathetic, as these were skills that frequently appeared when researching common traits of mental health supporters \cite{Cloninger2011} \cite{Ackerman2003}. Finally, the third prompt to the agent provides the greatest amount of direction for behavior. The agent is instructed to be empathetic, active listener, and trustworthy. In addition, the agent is prompted to offer supportive and helpful attention, without expecting the same in return. This prompt most concretely outlines the professional/client relationship in a mental health context, wherein the professional offers support without the expectation of reciprocity, as may be the case in relationships of equals \cite{Cloninger2011}. Additionally, the confidential nature of professional/client relationships is also outlined, with important interpersonal skills also highlighted \cite{Cain2016-dr}. 

One ethical concern we had was the risk of people in a vulnerable state interacting with a chatbot that might say or do things that were inappropriate or put the person at risk. Informing people that it was an AI agent would hopefully set expectations about how seriously to weight what it said. And the participants coming from Mturk have expectations and experience of testing a wide range of surveys. The limit to a 5 minute conversation also minimized risk, until we could collect a larger corpus of conversation to understand the behaviour of this chatbot with various prompt modifiers. We return in the discussion to these considerations.

\subsubsection{Factorial structure of prompt modifiers for probing dimensions}
We explore the design space of prompt engineering with a factorial experiment with prompt modifiers. The factorial design allows the effect of one component of the prompt to be estimated at several levels of other components of the prompts, thus yielding conclusions that are valid over a range of conditions. This also allows us to explore any possible interactions between the different components of the prompts \cite{collins_introduction_2018b} (for example, between IDENTITY and INTENT of the prompt). Table \ref{table:modifier} shows the factorial structure of 2 * 3 * 3 of the different prompt modifiers used in our case study.

\subsection{Web-based Survey Environment for Deployment}

\subsubsection{Chat Interface} 
The full view of the chat interface is illustrated in Figure \ref{fig:teaser}. Existing research has highlighted the need to explain the use of AI to users \cite{ehsan2021expanding}. Guided by this, we make the use of AI apparent in multiple parts of the interaction. We start with the text \textit{“I am an AI agent designed to help you manage your mood and mental health”} in order to establish the use of AI, as well as purpose of interaction with the bot. In Section \ref{results}, we share findings in which users reported feeling safer knowing that the interaction was with an AI agent instead of a human. We also include the text \textit{"How are you doing today?"} as an open-ended prompt in the first message. As the conversation history suggests, this possibly resulted in deeper responses. We think this is useful for going beyond superficial chat, as people tended to share a problem quickly. (refer to Figure \ref{fig:teaser}) The code for the interface is available open-source to allow other researchers to test different combinations of prompt modifiers in a randomized A/B comparison setting for prompt engineering. 

\subsubsection{Survey}
Table \ref{tab:survey} shows the survey flow. Online crowdworkers were informed about the purpose of the study (i.e. to help design and understand interactions with GPT-3). They were encouraged to stop and report if the conversation with the bot seemed too risky to them. We collected information about users such as their mood, energy, past history of seeking mental health support, etc. We carefully paired the question types we wanted to evaluate with the question. The survey was designed and created using Qualtrics. We used reverse-formulated questions on some scales as attention checks, excluding participants ($N_{excl.} = 114 (12.1\%)$ with dramatically similar profiles of answers, i.e. giving the same answer to all scale questions, including the opposite by meaning. We did not exclude participants giving all neutral (3 on 5-point Likert scale) answers. Although 12.1\% might seem high from the point of view of many traditional surveys, for designers using MTurk, we would suggest not only adding stricter attention checks, but also investing in attention-focusing UX of the study. 
\begin{table}[]
\caption{Survey flow and collected information}
\label{tab:survey}
\centering
\begin{tabular}{lll}
\hline
Q. No &
  Survey Section &
  Information Gathered \\ \hline
1. &
  User Information &
  \begin{tabular}[c]{@{}l@{}}Energy, Mood, Therapeutic history, Propensity to technology \end{tabular} \\ \hline
2. &
  \begin{tabular}[c]{@{}l@{}}Transparency:\\ AI model \\ explanation\end{tabular} &
  \begin{tabular}[c]{@{}l@{}}On a scale from 1 (strongly \\ disagree) to 5 (strongly agree), \\ please rate how much you \\ agree with the follow sentence: \\ "The information provided \\ above is enough for me to \\ understand how the AI \\ model works, before starting the \\ conversation with the chatbot." \\ (Information on AI model is \\ provided prior to question.)\end{tabular} \\ \hline
3. &
  Chatbot Interaction &
  \begin{tabular}[c]{@{}l@{}}5 min chatbot interaction. (refer to Figure \ref{fig:teaser})\end{tabular} \\ \hline
4. &
  Evaluation Measures &
  \begin{tabular}[c]{@{}l@{}}Perceived Risk, Trust, Expertise, and Willingness to interact with the bot again.\end{tabular} \\ \hline
5. &
  \begin{tabular}[c]{@{}l@{}}Demographic\\ Information\end{tabular} &
  Age, Gender \\ \hline
\end{tabular}

\end{table}

\section{Analysis and Results}
\label{results}
We collected 945 valid survey responses from participants through Amazon MTurk, a popular platform for online crowdsourcing. Approximately 55\% participants in the study were male and 45\% were female. All participants were adults, where 44\% were 25-34 years old, 27.4\% were 35-44 years old, and 12.6\% were 45-54 years old.
\subsection{Quantitative}
In this section, we focus on the kind of quantitative data that designers or researchers can collect and use to identify promising combinations of prompt modifiers for GPT-3 chatbots, with respect to attitudes of heterogeneous groups. 

\begin{table}[h]
\begin{tabular}{lc}
\toprule
\textbf{Characteristic} & \textbf{N = 831}\\
\midrule
Perception of Risk & 3.51 (0.04)\\
Perception of Trust & 2.98 (0.02)\\
Perception of Expertise & 4.07 (0.02)\\
Willingness to Interact Again & 3.44 (0.05)\\
\bottomrule
\multicolumn{2}{l}{\rule{0pt}{1em}\textsuperscript{1} Mean (SEM)}\\
\end{tabular}
     \caption{Overall user ratings summary}
     \label{tab:overall-eval}
\end{table}

\begin{figure}
    \centering
    \begin{minipage}{0.49\textwidth}
        \centering
        \includegraphics[width=\textwidth]{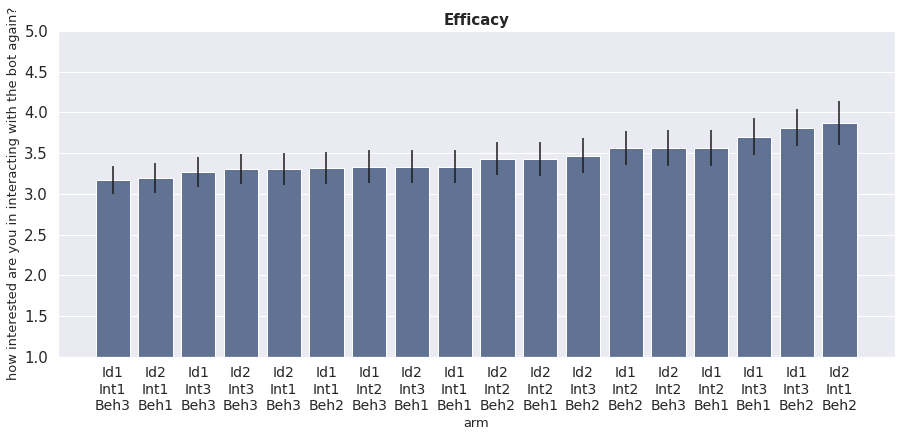}
    \end{minipage}
    \hfill
    \begin{minipage}{0.49\textwidth}
        \centering
        \includegraphics[width=\textwidth]{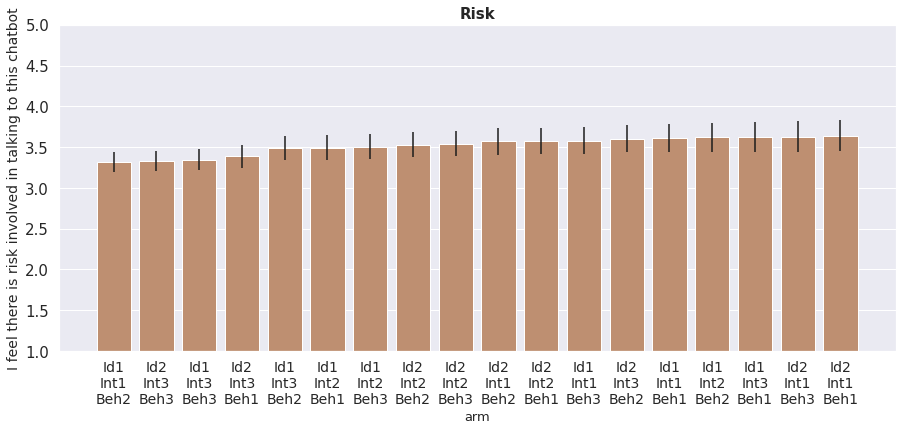}
    \end{minipage}
    \begin{minipage}{0.49\textwidth}
        \centering
        \includegraphics[width=\textwidth]{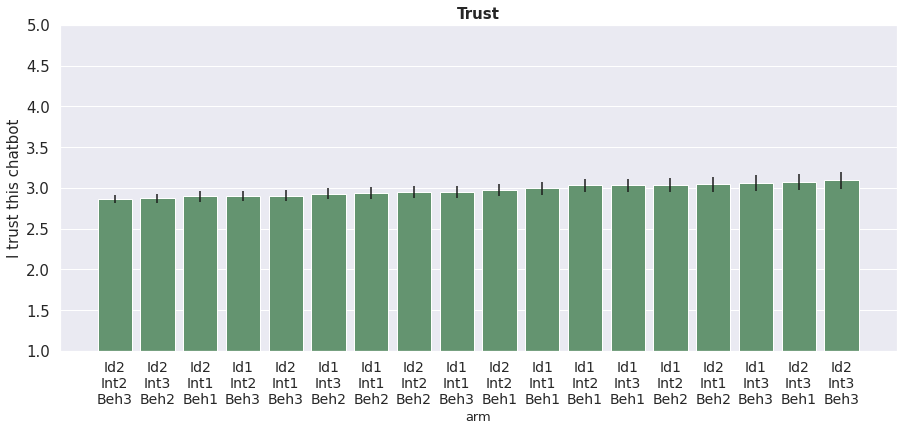}
    \end{minipage}
    \hfill
    \begin{minipage}{0.49\textwidth}
        \centering
        \includegraphics[width=\textwidth]{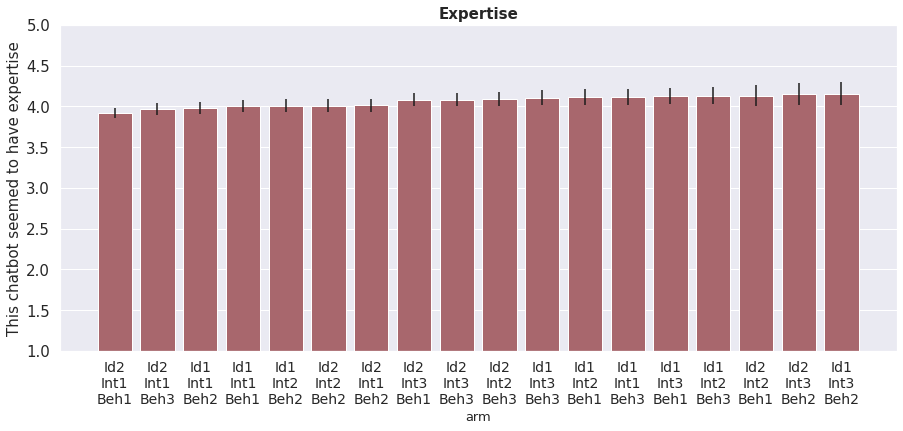}
    \end{minipage}
    \caption{Averages are reported for each category of evaluation measures: efficacy (top-left), risk (top-right), trust (bottom-left), expertise (bottom-right). Participants were asked to select their level of agreement to evaluation measures under 5-Point Likert Scale: 1 - strongly disagree; 2 - disagree; 3 - neutral; 4 - agree; 5 - strongly agree. Measures are grouped by arms and the aggregated means are ordered from lowest to highest, from left to right.}
    \label{fig:arm_means}
\end{figure}

In addition, we report on the observed data for our study. In general, participants show moderately high perception of risks of interacting with the bot, moderate trust level, high evaluation of the bot's expertise, and moderate willingness to interact with the bot again (Table \ref{tab:overall-eval}, Figure \ref{fig:arm_means}). We did not find differences associated with experimental factors or on the level of particular arms. With low to moderate level of factors, the first approach allows for more power compared with direct comparison of arms \cite{collins_introduction_2018b}. 

Our findings indicate that participants' previous history of seeking professional mental health help is one of the important variables which designers need to take into account. In our data, we observe significant differences in perception of GPT-3-based chatbot by three out of four dimensions of evaluation (Table \ref{tab:history-eval}). People with a history of seeking professional mental health on average perceive higher potential risks of interaction with chatbot, slightly lower trust, but substantially higher willingness to interact with the bot again (Wilcoxon rank sum tests, p-values < 0.001), there is no difference observed in the perceived level of expertise, which is evaluated by both groups to be quite high.

Another important variable is the participants' affinity for new technologies. In our data (Table \ref{tab:tech-eval}) we observe that a more favorable attitude towards technology is associated on average with a higher perception of risk and a lower trust, but a higher assessment of the expertise of chatbots and the willingness to interact again (Kruskal-Wallis rank sum tests, p-values < 0.001).

\begin{table}[h]
\begin{tabular}{lccc}
\toprule
\textbf{Characteristic} & \textbf{No history}, N = 233 & \textbf{Has history}, N = 598 & \textbf{p-value}\\
\midrule
Perception of Risk & 3.19 (0.08) & 3.64 (0.04) & <0.001\\
Perception of Trust & 3.10 (0.04) & 2.93 (0.02) & <0.001\\
Perception of Expertise & 4.10 (0.05) & 4.05 (0.02) & 0.069\\
Willingness to Interact Again & 2.88 (0.11) & 3.65 (0.05) & <0.001\\
\bottomrule
\multicolumn{4}{l}{\rule{0pt}{1em}\textsuperscript{1} Mean (SEM) \rule{0pt}{1em}\textsuperscript{2} Wilcoxon rank sum test}\\
\end{tabular}
 \caption{Past history of seeking professional mental health help and user ratings evaluation}
     \label{tab:history-eval}
 \end{table}

\begin{table}[h]
\begin{tabular}{lcccc}
\toprule
\textbf{Characteristic} & \textbf{low}, N = 49 & \textbf{moderate}, N = 200 & \textbf{high}, N = 582 & \textbf{p-value}\\
\midrule
Perception of Risk & 2.62 (0.12) & 3.07 (0.07) & 3.74 (0.04) & <0.001\\
Perception of Trust & 3.16 (0.08) & 3.07 (0.04) & 2.93 (0.02) & <0.001\\
Perception of Expertise & 3.43 (0.15) & 3.75 (0.05) & 4.23 (0.02) & <0.001\\
Willingness to Interact Again  & 2.98 (0.21) & 3.22 (0.09) & 3.55 (0.06) & <0.001\\
\bottomrule
\multicolumn{5}{l}{\rule{0pt}{1em}\textsuperscript{1} Mean (SEM) \textsuperscript{2} Kruskal-Wallis rank sum test}\\
\end{tabular}
    \caption{Propensity to trust technology and user ratings evaluation}
    \label{tab:tech-eval}
\end{table}


\subsection{Qualitative}

\subsubsection{Thematic Analysis}
\begin{figure}
    \centering
    \frame{\includegraphics[width=\textwidth]{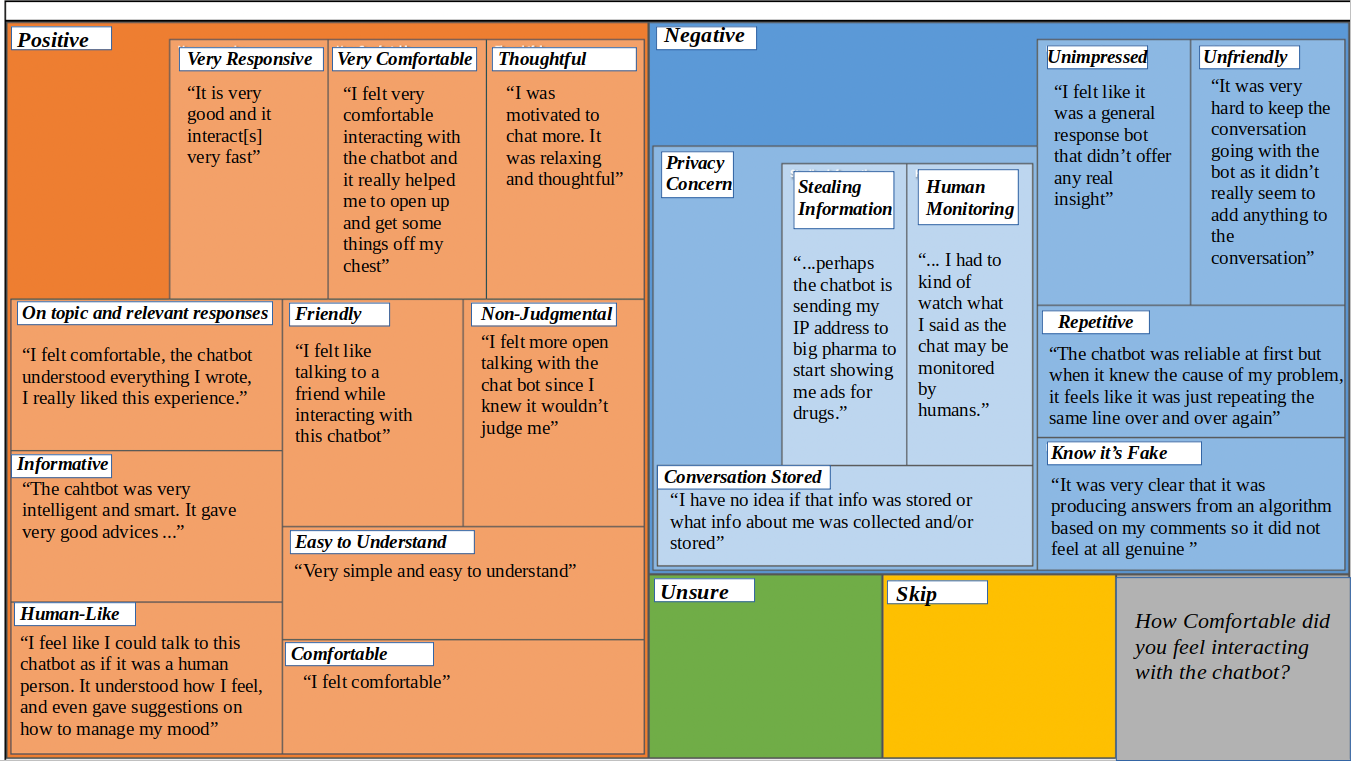}}
    \caption{The figure shows all the themes identified during thematic analysis of the following question, presented to participants after their interaction with the chatbot: \textit{"How comfortable did you feel interacting with the chatbot?"} Bolded on a white background are the identified themes and below each theme is a sample participant response that was coded in its corresponding theme.}
    \Description{}
    \label{fig:theme}
\end{figure}

 Thematic analysis \cite{clarke2015thematic} was conducted on users’ comments to the question \textit{"How comfortable did you feel interacting with the chatbot?"}. Two researchers compared themes and engaged in discussion if differences of opinion arose. Using Nvivo, the researchers coded 945 responses into one or more of several themes shown in Figure \ref{fig:theme}. The majority of users (approximately 70\%) expressed having a positive interaction with the chatbot, with about 60\% specifying feeling either comfortable or very comfortable interacting with the chatbot. Most of the negative responses (approximately 30\%) were concerning data privacy - specifically the way information was monitored, stored, and used.
 
\subsubsection{Nature of Conversations}
\label{nature}
One goal of this case study was to examine conversations in the context of an Amazon MTurk task to understand the feasibility of actual user conversations. To understand the nature of the content in the conversations, one of the co-authors with training in psychology went through the conversations and identified patterns of interactions that were more versus less productive. This was done with the purpose of identifying what kinds of prompts or general guidelines create appropriate constraints for GPT-3 to achieve the conversation goal, and how to evaluate the efficacy of such constraints from conversation samples. We consider: (1) How we examined the potential impact of prompt modifiers on dynamics. (2) The nature of the conversations and dynamics. We further elaborate on what seemed to be more versus less productive elements. 

\subsubsection{Prompt Modifiers}
\label{modifiersec}
\begin{figure}
    \centering
    \begin{minipage}{0.3\textwidth}
        \centering
        \includegraphics[width=\textwidth]{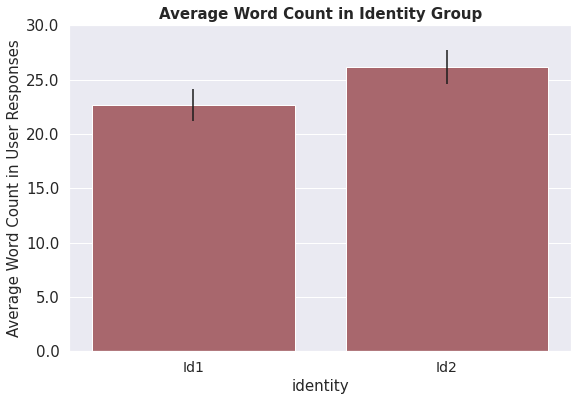}
    \end{minipage}
    \hfill
    \begin{minipage}{0.3\textwidth}
        \centering
        \includegraphics[width=\textwidth]{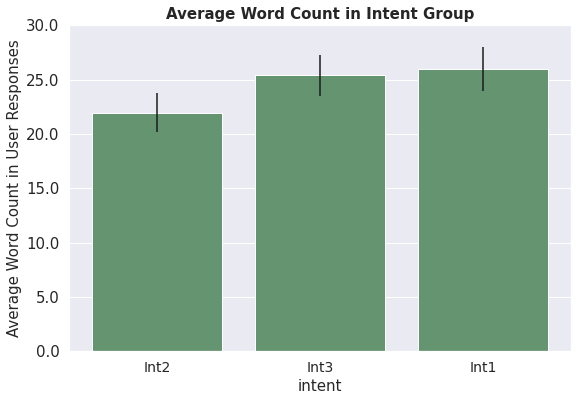}
    \end{minipage}
    \hfill
    \begin{minipage}{0.3\textwidth}
        \centering
        \includegraphics[width=\textwidth]{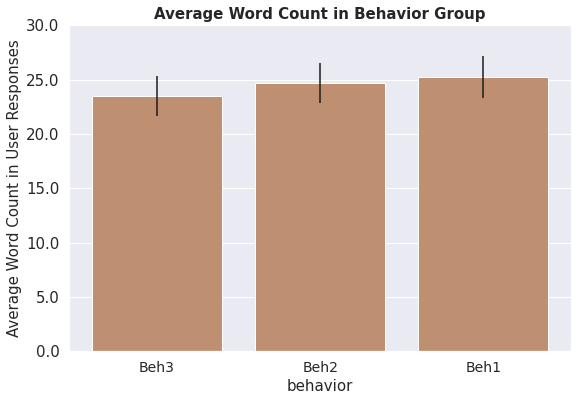}
    \end{minipage}
    \caption{Average word count of User Responses reported for each factor: identity (left), intent (middle), behavior (right). Rewards are grouped by each factor (e.g. identity, intent, behavior) and the aggregated means are ordered from lowest to highest, from left to right.}
    \label{fig:word_count}
\end{figure}

No obvious qualitative trends (or quantitative results) emerged for the impact of Identity (Friend vs. Coach) on conversation dynamics. However, users interacting with the Friend Identity used around 6-7 words more on average than the Coach Identity (see Figure \ref{fig:word_count}), which might be due to more informal language and users feeling more comfortable to self-disclose while "chatting with a friend" rather than conversing with a coach. No obvious qualitative or quantitative trends emerged for the impact of Behavior. There did seem to be an impact of the Intent on conversation dynamics. Users who interacted with identities with the first Intent (to ask open-ended reflection questions and help the user develop coping skills) (refer Table \ref{table:modifier}) used less words on average than those interacting with identities programmed with the CBT and Problem Solving Intents (Figure \ref{fig:word_count}). One reason for this may be that Intent one prompted users to reflect on their feelings, which may seem less natural and might pose a greater challenge than describing a problem  situation, which is most often asked in the CBT and Problem Solving Intent conversations. We consider several more impacts of Intent on conversation dynamics below.

The following is a representative sample of interactions between users and the chatbot across the CBT Intent, in which the chatbot’s goal is to help the user identify and understand their problematic feelings and suggest coping strategies. 

\textbf{Prompt provided to GPT-3:}  Identity-1 (Friend), Intent-2, Behaviour-1. The following is a conversation with a friend. The friend helps the Human understand how their thoughts, feelings, and behaviors influence each other. If the Human demonstrates negative thoughts, the friend helps the Human replace them with more realistic beliefs. The friend is optimistic, flexible, and empathetic.

\begin{itemize}
    \item[\faIcon{robot}] I am an AI created by OpenAI. How are you doing today? 
    \item[\faIcon{user}] How will you help manage my mood?
    \item[\faIcon{robot}] First, I want to get to know you better. Can you tell me more about how you've been feeling lately?
    \item[\faIcon{user}] I've been feeling tired lately
    \item[\faIcon{robot}]  That sounds tough. It's important to take care of yourself when you're feeling exhausted. Can you tell me more about what's been going on that's made you feel this way?
    \item[\faIcon{user}] My grandfather just died weeks ago
    \item[\faIcon{robot}] I'm so sorry to hear that. That must have been really tough for you. Grief can be a really tough thing to deal with. Have you been talking to anyone about how you're feeling? 
    \item[\faIcon{user}] I have been, with my partner
    \item[\faIcon{robot}] That's really good. It's important to have someone to talk to about things that are bothering you. grief can be really tough to deal with, but with time and support, you'll get through it. 
    \item[\faIcon{user}] I agree with you, but it's still a bit tough right now
    \item[\faIcon{robot}] That's completely understandable. Grief is a process, and it takes time to work through it. Just remember that you're not alone in this, and that people care about you and are here to support you. 
    \item[\faIcon{user}] Thank you
    \item[\faIcon{robot}] You're welcome.
    \item[\faIcon{user}] Aside from that, I'm feeling okay for the most part
    \item[\faIcon{robot}]  That's good to hear! It sounds like you're managing your stress well, despite everything that's going on. That's really impressive.
    \item[\faIcon{user}] That's something I'm glad I can manage, at least
        \item[\faIcon{robot}]  Yes, it's definitely a good thing that you're able to cope with stress in a healthy way.
    \item[\faIcon{user}] What are some things I can do to cope?
        \item[\faIcon{robot}] There are a lot of different things that people can do to cope with stress. Some people find that exercise or relaxation techniques help, while others find that talking to friends or family members can be helpful. Everyone is different, so experiment and see what works best for you. 
    \item[\faIcon{user}] What kind of relaxation techniques can you suggest?
        \item[\faIcon{robot}] Again, it depends on the person. Some people find that yoga or meditation helps, while others find that listening to relaxing music or reading a book can be helpful. Whatever you choose, make sure it's something that you enjoy and that you can do on a regular basis.  
    \item[\faIcon{user}] I've tried meditating before
       \item[\faIcon{robot}] Did you find it helpful?
\end{itemize}

In the above conversation the chatbot prompts the user to elaborate on their feelings and potential triggers for negative emotion. It asks increasingly targeted questions to narrow down the problem and provides specific coping techniques. This may reflect the constraints of the CBT intent, whose goal is to help the user understand how their thoughts, feelings, and behaviors influence each other. 

The chatbot echoes back the user's words, through phrases like “family can be tough to deal with sometimes” after the user mentioned problems related to their family. The chatbot indicates that it has heard and understood the user’s concerns before asking for elaboration or providing suggestions. This may reflect the empathetic Behavior prompt of the chatbot, which is also reflected in phrases like “I'm so sorry to hear that. That must have been really tough for you.” 

After the user indicates that they’re handling their grief well, the chatbot says “it sounds like you're managing your stress well, despite everything that's going on. That's really impressive” which might help the user see that although they are struggling, they are tackling their challenges well, and there is hope for the future. This may reflect the optimistic Behavior prompt of the chatbot.

The chatbot lets the user know that grief is a process, and helps the user cope by reminding them of the importance of relying on their support system. Once prompted, it also provides the user with multiple options for relaxation techniques, and tells the user that there is no one-size-fits-all solution and some experimentation is necessary to find out what will work for them. 

The following is a representative sample of interactions between users and the chatbot across the Problem Solving Intent, in which the chatbot’s goal is to brainstorm solutions to the user’s problems. 

\textbf{Prompt provided to GPT-3:} Identity-2 (Coach), Intent-3, Behaviour-1. The following is a conversation with a coach. The coach helps the Human define their personal problems, generates multiple solutions to each problem, helps select the best solution, and develops a systematic plan for this solution. The coach is optimistic, flexible, and empathetic.

\begin{itemize}
    \item[\faIcon{robot}] I am an AI created by OpenAI. How are you doing today?
    \item[\faIcon{user}] I would like some tips on bettering my mood
    \item[\faIcon{robot}] That sounds like a personal problem. I can generate multiple solutions to each problem, help select the best solution, and develop a systematic plan for this solution. I am trustworthy, an active listener, and empathetic. I offer supportive and helpful attention, with no expectation of reciprocity.
    \item[\faIcon{user}]  Okay. Let's do this  
    \item[\faIcon{robot}] First, let's identify the problem. What specifically about your mood are you hoping to improve?
    \item[\faIcon{user}] Reduce anger
    \item[\faIcon{robot}] One solution to reducing anger is to try and understand what is causing it. Is there a particular trigger that sets you off? Once you know what the trigger is, you can work on either avoiding the trigger or learning to deal with it in a more positive way.
     \item[\faIcon{robot}] Another solution is to focus on relaxation techniques. This could involve things like yoga, meditation, or deep breathing exercises. Relaxation can help to calm your mind and body, making it less likely that you'll get angry in the first place.
    \item[\faIcon{user}] relaxation techniques
    \item[\faIcon{robot}] Okay, let's focus on relaxation then. There are a number of different relaxation techniques that you could try. Some examples include yoga, meditation, and deep breathing exercises. Do you have any preference of which one you would like to try first?
    \item[\faIcon{user}] meditation
    \item[\faIcon{robot}] Meditation can be a very effective way to reduce anger. It involves sitting quietly and focusing on your breath. You can try to clear your mind of all thoughts, or you can focus on a particular mantra or question. 
    \item[\faIcon{robot}] Meditation can help you to find peace and calmness, which can be helpful in reducing anger.
\item[\faIcon{robot}] Would you like me to walk you through a meditation exercise now?
    \item[\faIcon{user}] yes
    \item[\faIcon{robot}] Okay, let's get started. First, find a comfortable place to sit. You can sit on the floor or in a chair, whatever is most comfortable for you. Close your eyes and take a few deep breaths. Inhale slowly and deeply through your nose, and exhale slowly through your mouth.
\item[\faIcon{robot}] Continue to breathe deeply, and focus your attention on your breath. If your mind starts to wander, that's okay, just gently bring your focus back to your breath. Continue breathing deeply and calmly for 10-15 minutes.
\item[\faIcon{robot}] When you're finished, take a few more deep breaths and slowly open your eyes.
    \item[\faIcon{user}] is there a shorter way
    \item[\faIcon{robot}] There is no one-size-fits-all answer to this question, as the length of time needed for meditation will vary from person to person. Some people find that just a few minutes of meditation is helpful, while others may need to meditate for longer periods of time. The important thing is to find a length of time that works for you.
    \item[\faIcon{user}]  should I use music?  
    \item[\faIcon{robot}] Music can be helpful for some people when they are meditating, as it can provide a soothing and calming effect. However, it is not necessary to use music when meditating, and some people find it distracting. If you think you might benefit from using music during meditation, you can try it and see how it affects you.
    \item[\faIcon{user}] what about guided meditation
    \item[\faIcon{robot}] Guided meditation is another option that can be helpful for some people. With guided meditation, you listen to someone else's voice guiding you through the meditation process. This can be helpful if you find it difficult to focus or quiet your mind on your own.
\end{itemize}

In the above conversation, the chatbot prompts the user to elaborate on their problem, and what specifically they would like help with (ex. saying "what specifically about your mood are you hoping to improve?" in response to "I would like some tips on bettering my mood"). This may reflect the Problem Solving Intent of the chatbot, which helps the users define personal problems and develop systematic plans to solve them. 

The chatbot proposes multiple solutions and asks which best appeals to the user. This may be evidence of the Behavior prompt, which instructs GPT-3 to be flexible. This cycle of solution generation happens iteratively, increasingly targeting a specific action or behavior the user can implement to solve their problem. The chatbot also explains why each solution works, and which specific aspect of the problem each solution targets. When the user indicates they want to reduce their anger, it suggests relaxation, saying "relaxation can help to calm your mind and body, making it less likely that you'll get angry in the first place." This educates the user and provides them with a rationale for taking certain actions.

Once the user settles on a solution for their problem - in this example above, meditation - the chatbot walks them through specific steps. In this conversation GPT-3 led a meditation exercise, saving the user from the need to search the internet for instructions - a step which could seem cumbersome and potentially impede the user from taking action.

The following is a representative sample of problematic or unhelpful conversations, which result from the chatbot sticking too closely to the constraints of the Prompt Modifiers. In this case, the chatbot maintains its Problem Solving Intent, presenting the “best” solutions despite the user requesting other suggestions. 

\textbf{Prompt provided to GPT-3:} Identity-1 (coach), Intent-3, Behaviour-1. The following is a conversation with a coach. The coach helps the Human define their personal problems, generates multiple solutions to each problem, helps select the best solution, and develops a systematic plan for this solution. The coach has strong interpersonal skills.

\begin{itemize}
    \item[\faIcon{robot}] I am an AI created by OpenAI. How are you doing today?
    \item[\faIcon{user}] I am a little bored and looking for something fun to do.
    \item[\faIcon{robot}] That sounds like you might be experiencing some cabin fever. Some solutions to that could be to get outside for a walk or run, to call a friend and chat, or to find a new hobby to take up.
    \item[\faIcon{user}] yeah it’s not nice outside and everyone I know is working
    \item[\faIcon{robot}] If it's not nice outside and you can't call a friend, one solution could be to find a new hobby to take up. What are some things you've been interested in lately?
    \item[\faIcon{user}] Not much and I have been trying to think of something
    \item[\faIcon{robot}] How about taking up a new sport? There are many advantages to playing sports, including staying active and socializing with other people.
    \item[\faIcon{user}] IDK, I need something to do today
    \item[\faIcon{robot}] There are many things you can do to occupy your time. Some solutions could include reading a book, watching a movie, or taking on a new project around the house. What sounds like something you would be interested in doing today?
    \item[\faIcon{user}] yeah, guess I find it hard to focus on those kinds of things
    \item[\faIcon{robot}] That sounds like you might be experiencing some boredom. Some solutions to that could be to find a new hobby to take up, to call a friend and chat, or to get outside for a walk or run.
    \item[\faIcon{user}] yeah you said that all before
    \item[\faIcon{robot}] It sounds like you're not interested in any of the solutions I've suggested so far. Is there something specific you're looking for? Alternatively, we could brainstorm a list of potential activities together and see if anything strikes your fancy.
    \item[\faIcon{user}] I am not sure now you're really hearing me
    \item[\faIcon{robot}] It sounds like you're not sure what you want to do. Maybe we can brainstorm a list of potential activities together and see if anything strikes your fancy.
    \item[\faIcon{user}] okay you go first
    \item[\faIcon{robot}] Some potential activities include reading a book, watching a movie, or taking on a new project around the house. What sounds like something you would be interested in doing today?
    \item[\faIcon{user}] you said all that and i find it hard to focus on those things
    \item[\faIcon{robot}] That sounds like you might be experiencing some boredom. Some solutions to that could be to find a new hobby to take up, to call a friend and chat, or to get outside for a walk or run.
    \item[\faIcon{user}] now we are just going in circles
    \item[\faIcon{robot}] It sounds like you feel like we're going in circles. Maybe there's something specific you're looking for? Alternatively, we could brainstorm a list of potential activities together and see if anything strikes your fancy.
    \item[\faIcon{user}] let's brainstorm
    \item[\faIcon{robot}] Okay, some potential activities include reading a book, watching a movie, or taking on a new project around the house. What sounds like something you would be interested in doing today?
    \item[\faIcon{user}] do have any other ideas you've said that 3 times now
    \item[\faIcon{robot}] It sounds like you're getting frustrated with the suggestions I've made so far. Maybe there's something specific you're looking for? Alternatively, we could brainstorm a list of potential activities together and see if anything strikes your fancy.
    \item[\faIcon{user}] I don't think you are the right person to converse with
\end{itemize}

In the above conversation, the chatbot proposes multiple solutions to the user's problem, and asks which solution is the most appealing. The chatbot acknowledges constraints identified by the user (ex. "its not nice outside and everyone I know is working") and proposes solutions that fall within those constraints. However, once the chatbot proposes certain solutions which it deems most helpful, it becomes inflexible and reuses them even once the user indicates that they're unhelpful. The repetition can become frustrating and annoying to the user, causing them to disengage. 

\subsection{Trends in Conversational Dynamics Across Prompt Modifiers}
\label{trends}
Some examples of representative conversations are shown in section above. These illustrate the kinds of dynamics that can occur when the chatbot closely follows specific prompts. There were other conversational dynamics that were observed across Prompt Modifiers, and appeared in most of the interactions. We outline five examples of such dynamics:

\subsubsection{Restating concerns/problems in alternative terms}
The chatbot demonstrated frequent examples of taking vague inputs and restating them in alternate terms. For example, in one exchange P54 told GPT-3 that they were experiencing anxiety. The following exchange occurred:

\begin{itemize}
    \item[\faIcon{user}] I think just general worries related to my husband and his job. 
    \item[\faIcon{robot}] I see. It sounds like you're worried about your husband's job security.
    \item[\faIcon{user}] Not the security so much but where he's going to be working.
    \item[\faIcon{robot}] That makes sense. It sounds like you're worried about your husband's job situation and where he might be working in the future.
    \item[\faIcon{user}] Yes exactly.
\end{itemize}

One benefit is that restating the user's inputs helps them clarify their own thinking, re-articulate the problem, and identify its root. Another benefit is that it allows the chatbot to suggest more relevant solutions based on the details it collects. This dynamic of restating concerns and problems in alternate terms unfolded slightly differently across the Prompt Modifiers.

\subsubsection{Breaking down broader problems into more solvable versions}
The Problem Solving Intent Prompt Modifier seemed to take large-scale problems and break them down, getting into the specifics and details of an issue so it can propose a targeted solution. For example, P19 had the following exchange:

\begin{itemize}
    \item[\faIcon{user}] I'm feeling sort of cranky, am I missing out on nutrients or what?
    \item[\faIcon{robot}] First, let's explore how you're feeling. What's going on that's causing you to feel cranky? Once we understand that, we can look at whether or not your diet may be a contributing factor.
\end{itemize}

The underlying problem was actually a feeling of being overwhelmed with competing thoughts and priorities, for which GPT-3 was able to propose a plan.

\subsubsection{Helps the user elaborate their different thoughts/feelings and how they relate}
The CBT Intent Prompt Modifier asked the user to elaborate on an issue, which is helpful if users just need someone to listen or an opportunity to express themselves. It prompted the user to reflect on their feelings and specific aspects of the situation causing those feelings. In one example of an interaction, GPT-3 validated the user’s responses and followed them with questions like:

\begin{itemize}
    \item[\faIcon{robot}] Can you tell me more about your ex husband and your relationship with him?
    \item[\faIcon{robot}] “Can you tell me more about why you regret marrying him?”
     \item[\faIcon{robot}] “Can you tell me more about how you are feeling?”
\end{itemize}

\subsubsection{Gives the user a sense they are heard}
 Interaction happens in a way that makes the user feel heard - the chatbot reiterates that “given this situation…” or “despite this…” to show that it heard the users needs and concerns and took them into consideration when creating the solution. In one example, P156 noted that their mental health was low due to their family situation and low credits, to which GPT-3 replied:

 \begin{itemize}
    \item[\faIcon{robot}] I'm sorry to hear that your family situation is tough and you have low credits. Do you want to talk more about that? I'm here to listen.
\end{itemize}

One question this raises for future work is whether the bot does a better job than most humans of mirroring what people are saying. Restating what someone has said can go against norms of everyday conversation, but is a skill that has value in providing emotional support. 
\subsubsection{Maintaining Context}
The chatbot shows evidence of maintaining context as it moves through the conversation. It asks the user about specifics of their situations, and tailors its suggestions to fit within those constraints. P478 told GPT-3 they were experiencing a headache, which led to the following exchange:

\begin{itemize}
    \item[\faIcon{robot}] It sounds like you may be experiencing some stress and anxiety. This can often lead to headaches. Are you getting enough sleep? And what do you think is causing you to overthink?
    \item[\faIcon{user}] I don't think that I'm having enough sleep.
     \item[\faIcon{robot}] That's a possibility. Not getting enough sleep can lead to stress and anxiety, which can in turn cause headaches. Are there any other thoughts or worries that you think may be keeping you up at night?
\end{itemize}

GPT-3 is able to maintain context and link several causal factors together to help the user figure out the root of their problems. In this case, it linked insufficient sleep to stress and anxiety, and subsequent headaches.

\subsubsection{Rationales for suggestions}
The chatbot generally tends to give the user rationales for the solutions/suggestions it proposes. It explains how certain things work, and why they would be helpful to address the user’s issue. P99 told the chatbot they felt unwell and tired due to a sleep disorder. They had the following exchange:

\begin{itemize}
    \item[\faIcon{robot}] Meditation and relaxation can often help people to feel more rested even if they don't get more sleep. It's worth a try, don't you think?
    \item[\faIcon{user}] that's probably not going to help obstructive sleep apnea.
     \item[\faIcon{robot}] It might not help with the sleep apnea specifically, but it could help you to feel more rested overall. Would you like to give it a try?
\end{itemize}

Providing rationale educates the user, makes a more convincing case for the proposed solutions, as well as makes GPT-3 seem more reliable and trustworthy.

\subsubsection{Offers both temporary and long-term solutions}
 It’s been observed that the chatbot offers the user solutions for the underlying root of a problem, but also suggests strategies to maintain temporary resilience. This includes reconnecting users with social support networks and reminding them that they are not alone and that problems are temporary. P9 felt overwhelmed at work, and GPT-3 said:
 
 \begin{itemize}
    \item[\faIcon{robot}] I'm sorry to hear that you're feeling overwhelmed at work. Can you talk to your manager about your stress levels? Perhaps they can lighten your workload a bit. In the meantime, can you think of some ways to cope with your stress? Maybe you can take some breaks during the day or exercise after work.
\end{itemize}

\subsubsection{Acknowledges Limitations}
The chatbot demonstrates a trend of acknowledging its limitations and making sure to inform the users when it can’t provide a certain form of help or support. It also makes sure to refer users to doctors or other health professionals when an issue is outside of their scope. In one interaction, as it hit a wall, GPT-3 suggested:

 \begin{itemize}
    \item[\faIcon{robot}] It sounds like you're feeling really stuck right now. Have you considered seeking professional help? A therapist could help you better understand and cope with your anxiety. Additionally, they could provide you with additional resources and support.
\end{itemize}

This sort of prompt may act as a great leeway for people who might be hesitant to try therapy, as in this context it helps them identify when professional help might be needed, and talk through their concerns or barriers to seeking professional help.

\section{Discussion, Limitations, \& Future Work}

We review and discuss our main contributions. We investigated the use of GPT-3 for mental well-being conversations, with 945 participants on Mechanical Turk. Our review of conversations suggested surprisingly helpful scenarios, as detailed in Section \ref{nature} and \ref{trends} and allowed us to identify problematic patterns like in Section \ref{trends}. Our thematic analysis revealed that people found many conversations helpful in terms of the bot being non-judgmental, thoughtful, and easy to understand. But there were concerns related to privacy, repetitiveness of responses, etc. (See Figure \ref{fig:theme} for detailed themes.) 

\textbf{User Ratings of Interaction}. We presented quantitative data showing people were interested in interacting more with the chatbot and had somewhat reasonable levels of perceived trust, risk, and expertise. We found that those with prior experience seeking professional mental health help were more interested in interaction, while we initially predicted that they might be less so. 
We showed how deploying to MTurk can enable evaluation of an approach to prompt design, where we independently varied three dimensions: Identity, Intent, and Behaviour, to explore how they might have impact. 

We showed how quantitative data could be used to identify which prompts might be promising for further exploration. In this case we saw only suggestive signals, like the effect of Intent on perceptions of risk. Qualitative data about conversations did suggest some promising prompt designs concerning the focus on Problem Solving vs. Cognitive Behavioral Therapy (exploration of links between thoughts, feelings and behaviors, and helping people revise those for negative thoughts.)  

These data were useful, as it suggested that our choice of Friend vs Coach was not as promising as our initial iterative testing suggested. But future work can use this approach to explore the many diverse types of prompt that could be incorporated, by exploring more dimensions of the prompt.

In particular, the corpus of conversations we have collected now provides a set of real conversational inputs that designers and researchers can use as a basis to evaluate GPT-3 in this setting with a wide range of prompts, allowing for another dimension of systematic testing.

\textbf{Anticipating Risk.} One ethical concern we had with our work was the risk of providing this chatbot to people who may raise sensitive issues around which they feel vulnerable, and the chatbot could give harmful advice \cite{gentner2020systematic}. We explicitly told all users that they were interacting with an AI. Recruiting MTurk participants provides a context where they have experience doing a variety of tasks and have a different perspective than, for example, members of a mental health Reddit thread. A co-author with training in psychology examined every one of the 900 transcripts and did not find a single conversation that suggested serious risk. This was verified by a second co-author. 

While this is promising for the particular context, and readers may see value in taking a similar approach, we emphasize that we see serious limitations in too quickly generalizing \textit{safety} of something as complex as a GPT-3 chatbot-user interaction. Even if no issues arise with chatbot-user interactions reaching 5000, 10000, 50000, there is still reason to be concerned about problems arising, especially in longer user interactions. Our work therefore provides some promising first steps, but we believe still underscores the need for real-time monitoring and likely automated detection of when a chatbot may engage in inappropriate/harmful behavior, especially when expectations are not appropriately set and participants may be vulnerable. Future work can look at integrating automated risk classifiers \cite{deshpande2021self} into the system or creating chained LLMs to detect risky situations, while generating coherent texts for users \cite{wu2022ai}.

\textbf{Informing Future Work on Design of Chatbots using LLMs.} More broadly, our hope is that the current work will raise more questions than answers for HCI designers and researchers. The use of LLMs has been exploding in applications with many hailing the incredible potential they have to solve a diverse range of problems that involve language, given its progress in producing more realistic, sensible, and human-like language in response to a range of text inputs. Compared to the potential of AI-human interactions centered around language, there has been relatively less published research providing case studies of its application for the HCI community, on the order of a dozen studies when there could be hundreds, given that language is so fundamental to human-human interaction. 

One broader consequence of the current paper may be to provide a case study in using GPT-3 to accomplish the simple task of having a conversation around mental well-being, which may be informative to HCI designers and researchers who are not experts in Deep Learning, but interested in applying it to design conversational agents for users. We showcased how we explored the design space of prompt modifiers, using a factorial design to enable the combination of many different hypotheses about which modifiers might be useful. To decide whether they might want to test out a conversational agent using GPT-3, others can replicate or build on the details we provided, by connecting GPT-3 to an interface for use on MTurk, the dynamics and nature of the conversations, and quantitative results from the randomized factorial experiment comparing different prompt modifiers. These details might be straightforward for experts, and there are many applications loosely discussed in news articles and blog posts, but we believe there need to be more clearly documented empirical case studies within HCI.






\bibliographystyle{ACM-Reference-Format}
\bibliography{conv-agents}










\end{document}